# An alternative simulation approach for surface flashover in vacuum using a 1D2V continuum and kinetic model

Guang-Yu Sun[1], Ru-Hui Lian[1], Shu Zhang[1], Xiong Yang[1], Muhammad Farasat Abbas[2], Chao Wang[1], Bao-Hong Guo[3], Bai-Peng Song[1] and Guan-Jun Zhang[1]

[1] State Key Laboratory of Electrical Insulation and Power Equipment, School of Electrical Engineering,
Xi'an Jiaotong University, Xi'an, Shaanxi 710049, People's Republic of China
[2] National University of Sciences and Technology (NUST), Islamabad, 44000, Pakistan
[3] Centrum Wiskunde and Informatica (CWI), Amsterdam, Netherlands

E-mail: gjzhang@xjtu.edu.cn, bpsong@xjtu.edu.cn

**Abstract**

Surface flashover across insulator in vacuum is a destructive plasma discharge which undermines the behaviors of a range of applications in electrical engineering, particle physics, space engineering, etc. This phenomenon is widely modeled by the particle-in-cell (PIC) simulation, here the continuum and kinetic simulation method is first proposed and implemented as an alternative solution for flashover modeling, aiming for the prevention of the unfavorable particle noises in PIC models. A 1D2V (one dimension in space, two dimensions in velocity) kinetic simulation model is constructed. Modeling setup, physical assumptions, and simulation algorithm are presented in detail, and a comparison with the well-known secondary electron emission avalanche (SEEA) analytical expression and existing PIC simulation is made. Obtained kinetic simulation results are consistent with the analytical prediction, and feature noise-free data of surface charge density as well as fluxes of primary and secondary electrons. Discrepancies between the two simulation models and analytical predictions are explained. The code is convenient for updating to include additional physical processes, and possible implementations of outgassing and plasma species for final breakdown stage are discussed. The proposed continuum and kinetic approach is expected to inspire future modeling studies for the flashover mechanism and mitigation.

## 1. Introduction

The surface flashover in vacuum is a plasma breakdown which occurs across an insulator under high applied voltage. Flashover is accompanied by insulation failure and therefore jeopardizes the safe operation of pertinent devices. Hence improved understanding of the flashover phenomenon and its mitigation is expected. Surface flashover is found in high voltage transmission apparatus, spacecraft, pulsed-power devices, particle accelerator, etc. [1-11]. Depending on the background pressure and gas species, plasma discharge during surface breakdown varies from streamer discharge, corona discharge, to Townsend-like discharge, etc. [8, 12, 13] In high degree of vacuum without ionization source for discharge initiation, flashover begins from the cathode triple junction (CTJ) where electrode, insulator, and vacuum adjoin [14]. Due to the transient and nonequilibrium nature of the flashover process, numerical modeling provides valuable supplements to the experimental observations [15-22]. The particle-in-cell (PIC) method is a widely adopted numerical approach that is implemented in most if not all recent vacuum flashover simulation studies. The present work aims for providing an approach other than the particle model that may serve as an alternative solution for surface flashover study, i.e. the continuum and kinetic model.

Modeling of surface flashover in vacuum is complicated due to various physical processes involved. According to the secondary electron emission avalanche (SEEA) theory, flashover is initiated by the field emission at CTJ due to local electric field distortion. The emitted electrons are accelerated by the applied field and induce secondary electron emission (SEE) on the insulator surface. An avalanche occurs when the created secondary electron (SE) continues to produce more SEs while charging the insulator surface positively [23]. A final breakdown is only possible after consistent electron collisions with insulator unleash neutrals that were previously adsorbed in the insulator, and a final plasma breakdown within the local high-pressure region occurs [24]. The time scale, particle species, and underlying physics are drastically distinct among different flashover stages, and systematical simulation of the complete flashover process is challenging.

Currently employed PIC models have been widely implemented in flashover study, covering physical mechanism of flashover process in separate stages [16, 25-31], and flashover mitigation using surface microstructure [32-35], surface processing [36], external electromagnetic field, etc. [37-39]. In order to improve the simulation precision, the number of macroparticles in PIC simulation must be sufficiently large, which inevitably enhances the computational cost and is usually accompanied by considerable numerical noises due to particle discretization.

In addition, strong fluctuations of the physical quantities even when SEEA reaches saturation (without outgassing) were observed in previous PIC simulations [40-42]. It is unclear whether these are realistic waves/instabilities or purely numerical. Recent simulation of SEEA including the outgassing process suggested that higher harmonics can be generated by beam-plasma interactions [24], but it remains elusive whether realistic waves exist in the SEEA development stage without outgassing. The present work aims at deciphering this issue by using an alternative simulation approach featuring significantly lower numerical noises.

Though never applied in surface flashover simulation, the continuum and kinetic simulation approach has been shown to be a supplementary solution for PIC model for a range of plasma conditions, and features some advantages if applied appropriately. The fact that kinetic simulation speed is not sensitive to particle number, and its noise-free data enables its implementation in a variety of industrial plasma simulations, such as plasma-surface interactions [43], plasma transport [44], capacitively coupled plasma [45, 46], multipactor [47], glow discharge [48], in addition to fusion plasma as well as astrophysical plasma simulations [49-52]. In the present work, we propose for the first time to apply a 1D2V (one dimension in space, two dimensions in velocity) continuum and kinetic simulation in surface flashover modeling to add more insights into improved flashover modeling technics. The reduced model with only one spatial dimension perpendicular to the insulator surface allows to probe the flashover development with significantly lower computational cost while retaining the salient physical details during flashover. Such reduced model has been widely adopted with PIC simulation approach though not yet with the continuum and kinetic approach [53-55]. Additionally, a comparison of the kinetic simulation results with the existing SEEA theory is expected to reveal more detailed underlying SEEA properties that are not included in the theoretical model.

The article structure is as follows. In section 2, mechanism, numerical scheme, algorithm, specific treatment for surface flashover, and choices of simulation parameters are presented in detail, for the constructed 1D2V continuum and kinetic model. In section 3, the obtained simulation results including surface charge density and surface electron fluxes are shown and a comparison with SEEA analytical prediction is made. In section 4, the kinetic model is compared with the existing 2D PIC model and the pros and cons of the kinetic model are analyzed. Section 5 discusses possible future code upgrades to include the complete flashover process. Concluding remarks are given in section 6.

## 2. Model setup

In this section, the kinetic simulation setup, modeling assumptions and algorithms in the surface flashover background are presented in detail. The kinetic simulation, different from the PIC model that simulates the movement of individual super-particles, calculates the evolution of particle velocity distribution function (VDF) according to the following Boltzmann kinetic equation.

$$\frac{\partial f(x,v)}{\partial t} + \boldsymbol{v} \cdot \nabla f + \frac{q[\boldsymbol{E}+\boldsymbol{v}\times\boldsymbol{B}]}{m} \nabla_v f = \frac{\partial f}{\partial t}\Big|_{coll} \quad (1)$$

Here position $\boldsymbol{x}$, velocity $\boldsymbol{v}$, electric and magnetic field $\boldsymbol{E}$ and $\boldsymbol{B}$ are all 3D vectors. $q$ and $m$ are charge and mass of the species, and $f$ is the velocity distribution function. RHS is the VDF source or sink due to interparticle collisions or external factors. Each species corresponds to a kinetic equation, and all kinetic equations are coupled by Maxwell equation or Poisson equation, depending on the context of study. Equation (1) is reduced to the Vlasov equation if no collision term exists. Physical quantities such as density, particle and heat flux, pressure are derived from the VDF at given time and location, usually through numerical integral. A complete kinetic simulation technically requires six dimensions in total, three in space and three in velocity, i.e. 3D3V. In reality, such complete simulation is rarely performed due to high computational cost, and pertinent physical assumptions are adopted to reduce the dimensions needed. For example, kinetic simulation in magnetized plasma can reduce the required dimension by one, through averaging over the gyro-phase angle, also called gyrokinetic simulation [50]. When studying unmagnetized sheath, 1D1V kinetic model is commonly adopted which only considers space and velocity in direction perpendicular to the surface, assuming parallel plate geometry [43, 56]. 1D1V kinetic model was also adopted for double-surface multipactor simulation in the waveguide with high-power microwave [47].

For surface flashover modeling, 2D simulation is widely employed [15, 16, 22, 27, 41], which considers space dimensions in direction perpendicular to insulator ($x$), and the direction perpendicular to electrode ($y$). A schematic of the considered insulator system is shown in Figure 1, where the insulator bridges cathode and anode, with electron emission from CTJ and SEEA developing across the insulator surface. Here it is assumed that the physics in direction parallel to both dielectric and insulator ($z$ direction, pointing inward the paper) is homogenous, which is valid only if the applied parallel electric field ($E_y$) is uniform in $z$ direction. This assumption is naturally valid for a cylindrical/conical insulator. For other electrodes, the assumption is justified near the central $y$ axis unless the electrode has extremely small curvature radius, e.g. needle electrode. The assumption is widely adopted in existing modeling works and is shown to exhibit good consistency with the practical insulation systems.

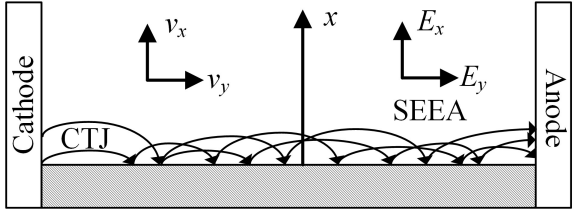

Figure 1. Schematic of the considered surface flashover model.

A further model simplification is possible by disregarding the spatial dimension $y$, and only considers the velocity in this direction. This is because during the development for SEEA along $y$ direction, the charging dynamics of a given $y$ position is the same as a position having $\Delta y$ distance away, except with a time difference of $\Delta t = \Delta y / v_{SEEA}$, with $v_{SEEA}$ the propagation velocity of the SEEA. It is then possible to focus on one single location and construct a 1D2V model. This will be further discussed later in this section.

All locations (except cathode adjacency) on the insulator surface have the same surface charge density after the SEEA has fully developed and covers the whole insulator surface, which is also called saturated secondary electron emission (SSEE) stage [57]. In SSEE, the average secondary electron emission yield (SEEY) is one, meaning that one newly created secondary electron, after being accelerated by the parallel electric field $E_y$, always induces another SE upon collision with insulator. The assumption of SSEE stage was also employed in the single-particle analyses of SEEA process originally developed by Boersch [23]. The considered dimensions ($x$, $v_x$, $v_y$) in the simulation model are as shown in Figure 1.

In the present 1D2V model, Equation (1) is simplified into the following form. Since the plasma current is not strong enough to create remarkable magnetic field and no external magnetic is applied, the magnetic force term is neglected.

$$\frac{\partial f_e(x,v_x,v_y)}{\partial t} = -v_x \frac{\partial f_e}{\partial x} - \frac{q_e E_x}{m_e}\frac{\partial f_e}{\partial v_x} - \frac{q_e E_y}{m_e}\frac{\partial f_e}{\partial v_y} + \frac{\partial f_e}{\partial t}\bigg|_{ee} + S_{CTJ} \quad (2)$$

The subscript $e$ represents electron, and the last two terms in RHS represent electron-electron Coulomb collision and the electron source due to CTJ field emission, to be expatiated later. Since the present work focuses on the SEEA process without outgassing and subsequent plasma discharge, electron is the only considered species. Upgrades including more plasma species will be discussed in section 5.

The simulation begins with zero space and surface charge everywhere in the simulation domain, corresponding to an experimental condition with high degree of vacuum (below $10^{-4}$ Pa). At each time step, advections, collision and source terms in RHS of Equation (2) are performed numerically in explicit, upwind scheme to update the 3D matrix $f_e$.

For the two velocity advections, electric field components $E_x$ and $E_y$ are required. Since there's no contribution of surface charge to $E_y$, as opposed to the 2D model [16, 17], $E_y$ is only determined by the applied electric field between the electrodes. In the simulation, $E_y$ is given as a constant input parameter and is always negative with cathode on the left, as illustrated in Figure 1. $E_x$ is determined by a 1D Poisson equation solver with Neumann boundary condition on dielectric boundary and Dirichlet boundary condition on vacuum boundary, with $E_{x,dielectric} = \sigma_e / 2\varepsilon_0$ and $V_{x,vacuum} = 0$. $\varepsilon_0$ is vacuum permittivity and $\sigma_e$ is surface charge density. Note that the effects of surface charge on spatial electric field distribution is complex which is closely linked with the subsurface charge trapping, migration and dissipation [10, 11, 58-61], and the boundary condition here is merely a simplification in 1D condition considering only above-surface processes.

Electron-electron Coulomb collision is characterized by the BGK collision operator[62] with frequency $v_{ee}$, expressed as follows:

$$\frac{\partial f_e}{\partial t}\bigg|_{ee} = v_{ee}(n_e \tilde{f}_{e0} - f_e) \quad (3)$$

$$v_{ee} = \frac{n_e e^4 \ln(\Lambda)}{12\pi^{1.5}\varepsilon_0^2 \sqrt{m_e} T_e^{1.5}} \quad (4)$$

Here the Coulomb logarithm $\ln(\Lambda) = 10$, $n_e$ is electron density at position $x$, $\tilde{f}_{e0}$ is the normalized electron VDF in equilibrium with electron temperature $T_e$, consisting two velocity components:

$$\tilde{f}_{e0}(x,v_x,v_y) = \frac{m_s}{2\pi T_e}\exp(-\frac{m_e v_x^2 + m_e v_y^2}{2T_e}) \quad (5)$$

Equation (3) removes a fraction of electron VDF and replenishes it with VDF in equilibrium.

Since there's no electron in the entire domain in the beginning of simulation, electron source term is required to initiate and sustain the discharge. Here the source electrons are provided by field emission from CTJ. The field emission current density $J_{CTJ}$ is calculated by the Fowler-Nordheim formula [63]:

$$J_{CTJ} = \frac{e^3}{8\pi h \varphi_m}E_{CTJ}^2\exp(-\frac{8\pi\sqrt{2m_e\varphi_m^3}}{3ehE_{CTJ}}) \quad (6)$$

Here $h$ is Planck's constant, $e\varphi_m$ is the work function of electrode material, and $E_{CTJ}$ is the CTJ electric field. $E_{CTJ}$ is taken as $|E_y|$ and the field enhancement effect is not considered. The CTJ source is located between 10-15μm away from the insulator and is uniformly distributed in the following form:

$$S_{CTJ} = \frac{J_{CTJ}}{eh_{CTJ}}\tilde{f}_{e0,FE} \quad (7)$$

Here $h_{CTJ}$ is the width of CTJ emission region (5μm) and $\tilde{f}_{e0,FE}$ is the normalized VDF of electrons created by field emission. The exact form of $\tilde{f}_{e0,FE}$ is complicated [64], and here a simplification is made to consider $\tilde{f}_{e0,FE}$ as a half Maxwellian with temperature $T_{fe}$ in $v_y$ direction and full Maxwellian in $v_x$ direction, similar to Equation (5). $T_{fe}$ is determined by Fermi level of cathode material. The simulation results are however not sensitive to the initial field-emitted electron VDF, as it is the parallel field acceleration that primarily drives the source electrons.

To form the SEEA, secondary electron emission on insulator is indispensable. The treatment of SEE process here is different from the previously used averaged SEEY for all electrons in emissive sheath studies [56, 65], instead a SEEY matrix $\delta_e$ of dimension $d_{vx}\times d_{vy}$ is constructed, with $d_{vx}$ and $d_{vy}$ the grid number in $v_x$ and $v_y$ direction. The matrix $\delta_e$ represents SEEY of each element $(v_{x,i},v_{y,j})$ in phase space. Each element of the matrix is calculated by the following empirical formula[66]:

$$\delta_e = 1.526\delta_{max}\left(1+\frac{k_s\alpha_{in}^2}{2\pi}\right)\frac{1-\exp(-z^{1.725})}{z^{0.725}} \quad (8)$$

$$z = 1.284A_{in}/[A_{max}(1+\frac{k_s\alpha_{in}^2}{\pi})] \quad (9)$$

Here $A_{max}$ and $\delta_{max}$ are two parameters characterizing the SEEY curve of a dielectric material, which are the maximum SEEY and the corresponding normal incident energy in the curve. $k_s$ is the smoothness factor and here the normal surface condition $k_s = 1$ is chosen. $A_{in}$ is the electron incident energy in eV and $\alpha_{in}$ is the incident angle with respect to the direction perpendicular to the insulator:

$$A_{in}(v_{xi},v_{yj}) = 0.5m_e(v_{x,i}^2+v_{y,j}^2)/e \quad (10)$$

$$\alpha_{in}(v_{xi},v_{yj}) = \mathrm{atan}(\left|\frac{v_{y,j}}{v_{x,i}}\right|) \quad (11)$$

Note that SEE occurs at the surface only when $v_x < 0$, here $v_{x,i}$ and $v_{y,j}$ are grid point velocities with $1\leq i\leq dvx$ and $1\leq j\leq dvy$.

The SEEY matrix is determined in the beginning of simulation and is reused in each iteration. Individual SEE flux is calculated based on $\delta_e$ and the VDF at dielectric boundary ($x = 0$), which is summed up for $v_x<0$ and all $v_y$ to get the total SEE flux:

$$\Delta\Gamma_{see}(i,j) = \delta_e(i,j)\Delta\Gamma_{pe} \quad (12)$$
$$\Delta\Gamma_{pe} = -v_{x,i}f_e(1,i,j)dv_xdv_y \quad (13)$$
$$\Gamma_{see} = \sum_{v_{x,i}<0}\sum_{v_{y,j}}\Delta\Gamma_{see}(i,j) \quad (14)$$

Here $\Delta\Gamma_{pe}$ is the primary electron flux element. The index 1 in $f_e$ of Equation (13) represents the dielectric boundary, and the negative sign is because $v_x<0$.

With the total SEE flux to emit from the dielectric surface, the boundary condition for $f_e$ at the dielectric surface is set as:

$$f_e|_{x=0,v_x>0} = \frac{\Gamma_{see}}{\sqrt{2\pi}}\left(\frac{m_e}{T_{se}}\right)^{1.5}\exp(-\frac{m_ev_x^2+m_ev_y^2}{2T_{se}})(15)$$

The secondary electrons VDF is assumed half-Maxwellian in $v_x$ direction and full Maxwellian in $v_y$ direction, with temperature $T_{se}$. Equation (15) is obtained by the following flux definition:

$$\Gamma_{see} = \int_{-\infty}^{+\infty}\int_{0}^{+\infty}f_{e,x=0}v_xdv_x\,dv_y \quad (16)$$

For vacuum boudnary located at $x = h_{max}$, all electrons passing through the boudnary are fully absorbed, with:

$$f_e|_{x=h_{max},v_x<0} = 0 \quad (17)$$

In the present simulation, only SEE is considered and both elastic and inelastic reflection can be included by introducing a reflection coefficient $R_f$, such that electrons have $1-R_f$ probability to induce SEE, whose treatment is introduced above. Several different algorithms to implement the electron reflection were introduced in a recent literature [65].

When calculating $\Delta\Gamma_{see}(i,j)$, the surface charge density $\sigma_e$ is also updated. The change of surface charge density in each time step is:

$$\Delta\sigma_e = e\Delta t\sum_{v_{x,i}<0}\sum_{v_{y,j}}[\delta_e(i,j)-1]\Delta\Gamma_{pe} \quad (18)$$

Here $\Delta t$ is the time step. The updated surface charge density is then used to set $E_x$ boundary condition for the next time step.

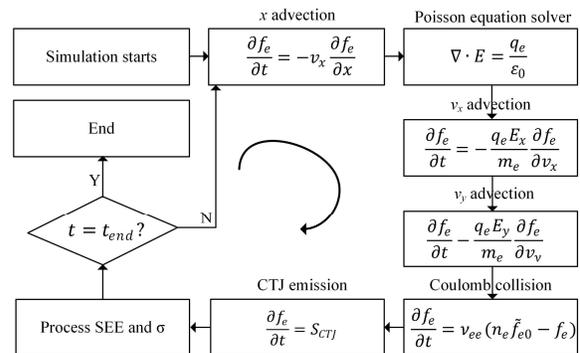

Figure 2. Flow chart of the program execution.

A flow chart of the program execution is shown in Figure 2, summarizing abovementioned modeling procedures, and the convergence is achieved if the surface charge density and surface flux change within 0.1% within $10^4$ time steps.

Note that above simulation setup aims at simulating the SEEA formation near the cathode triple junction. According to the SEEA theory, the SEEA develops from the cathode to anode and gradually covers the entire insulator surface. During the SEEA expansion, region covered by SEEA achieves saturation with a constant surface charge density and SEEY of one, such that all incident electrons from the cathode side are converted in the form of SE to the uncovered region with the same number of electrons, as if the incident electrons arriving at the uncovered region come directly from the CTJ. It is hence assumed in the classic SEEA theory that all SEEA dynamics are homogeneous in the $y$ direction [67], such that the present simulation results are actually applied to any arbitrary position of the $y$ direction.

Default simulation parameters are given as follows and are unchanged unless specified. Electron temperatures $T_e = 2eV$, $T_{fe} = 4eV$, $T_{se} = 2eV$, $E_y = -5 \times 10^6 V/m$, $\varphi_m = 4.08V$, $h_{max} = 20\mu m$. SEE coefficients are $\delta_{max} = 6.5$, $A_{max} = 650eV$, resembling alumina ceramics. Grid number is 200 in $x$, 401 in $v_x$ and 201 in $v_y$. Velocity range is 16 times the thermal velocity $v_{Te} = \sqrt{T_e/m_e}$ in $v_x$ and 40 times the thermal velocity in $v_y$. Time step $\Delta t = 5 \times 10^{-15}s$. Grid and time resolutions are prescribed by the CFL conditions. The chosen parameters are based on previous PIC modeling works of surface flashover in vacuum [15-19, 27, 41].

## 3. Simulation results and comparison with theory

In this section, results of 1D2V continuum and kinetic simulation are presented, focusing on the surface and space charging behaviors during the SEEA development. A comparison with existing SEEA analytical expression is made to validate the simulation code.

Electrons released by field emission at CTJ are strongly accelerated by the applied field $E_y$, and a fraction of the field emission electrons with negative initial $v_x$ component will collide with the insulator surface, inducing secondary electron emission. Note that under the intense applied field ($\sim 10^5$ V/m), incident primary electron can carry energy of decades to over one hundred eV. Therefore, more than one SEs are unleashed and the insulator surface is positively charged, i.e. secondary electron flux is above the primary electron flux, shown in simulation results of Figure 3.

In the beginning of simulation, no electron exists in the entire domain and it takes some time for source electrons from CTJ to arrive at the dielectric surface, Figure 3(b). Once electrons arrive at the surface, strongly accelerated electrons charge the surface positively, which further attracts electrons, causing a "spike" of electron fluxes in the first 5 ps. Electron fluxes increase slower as the electron space-charge effect gradually becomes obvious, which shields the surface charge field. The primary and secondary electron fluxes eventually become equal, when the surface charge density reaches saturation.

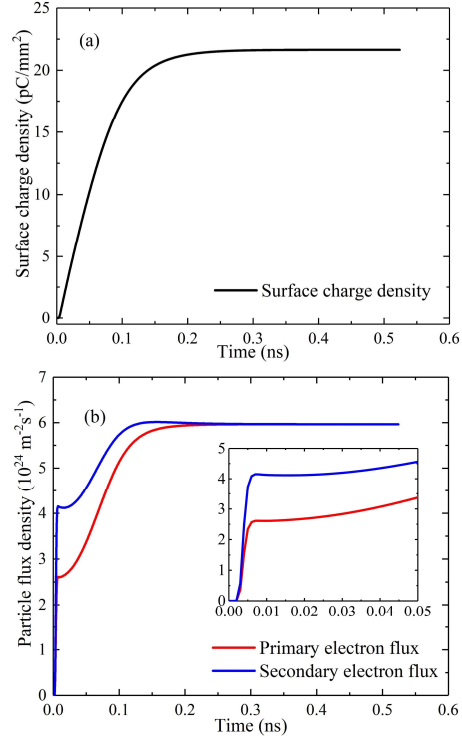

Figure 3. Time evolution of (a) surface charge density and (b) primary and secondary electron fluxes during SEEA formation. A subplot is given in (b) to show the change of electron fluxes in the first 0.05 ns.

The saturation of SEEA is achieved in the following way. As positive surface charges build up, newly emitted SEs are attracted back to surface. Higher surface charge density $\sigma_e$ decreases the electron time of flight and the incident energy upon electron collision with the insulator, as the perpendicular field $E_x$ is proportional to the surface charge density. Eventually, a balance is achieved when one emitted SE creates another one SE after being accelerated by $E_y$, combined with a saturated surface charge density and balanced primary and secondary electron fluxes, achieved at approximately 0.3 ns in Figure 3.

Above processes can be depicted by an analytical model based on single-particle analyses, i.e. the SEEA analytical expression [23]. For a SE emitted in direction perpendicular to the insulator surface, its time of flight is $t_{ef} = 2\sqrt{2A_0/m_e}/|a_{ex}|$ and the traveled distance $l_{ef} = 0.5 a_{ey} t_{ef}^2$, with the initial energy $A_0$ and the vertical and parallel acceleration $a_{ex}$ and $a_{ey}$. For a fully developed SEEA, the electron energy upon arriving at insulator again must be $A_1$, the required energy to produce one new SE. Therefore, the following equation must be satisfied [23]:

$$\frac{2}{m_e}(A_1 - A_0) = 2a_{ey}l_{ef} \quad (19)$$

Considering cosine distribution as the initial SE's angular distribution, Equation (19) is rewritten as:

$$\sigma_e = 2\varepsilon_0 |E_y| \left[0.5\left(\frac{A_1}{A_0} - 1\right)\right]^{-0.5} \quad (20)$$

Note that here the boundary condition $E_x = \sigma_e/(2\varepsilon_0)$ and the relation $a_{ex}/a_{ey} = E_x/E_y$ is used. The adopted theory has several key assumptions that may not be valid in a range of conditions, hence leading to discrepancies when compared with the simulation results, to be discussed below.

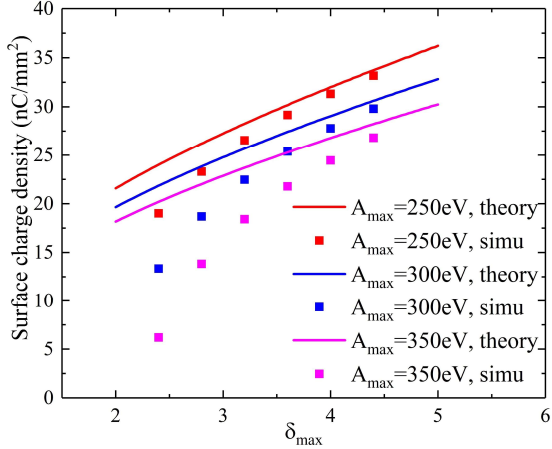

Figure 4. Comparison of simulated surface charge density with SEEA analytical prediction for a range of $A_{max}$ and $\delta_{max}$ values.

A scan of SEEY parameters $A_{max}$ and $\delta_{max}$ is performed and the obtained simulation results are compared with the SEEA theory. The simulation in general predicts lower $\sigma_e$ than the SEEA theory, and the discrepancy is lower at high $\sigma_e$ levels. With higher $\delta_{max}$ and lower $A_{max}$, surface charge density increases. This is because the factor $A_1$ can be approximated by $A_1 \sim \delta_{max}/A_{max}$, and $\sigma_e$ increases with $A_1$.

The discrepancy between theory and simulation mainly consists of four sources. First is the different adopted angular distributions. The theory assumes the cosine distribution where the differential SE flux of solid angle $d\Omega$ is proportional to the emission angle with respect to the insulator surface normal. The cosine distribution is a 3V distribution that cannot be self-consistently implemented into the present 2V framework. In the simulation, a 2V Maxwellian is chosen for simplicity. The second source of discrepancy comes from the theory assumption that all emitted SEs have the same initial energy $A_0$, instead of the Maxwellian distribution employed in the simulation. The third reason is that the theory doesn't consider incident angle. Note that in reality, the electrons collide on insulator with grazing angle. The last discrepancy is due to theory limit at low surface emissivity. The SEEA theory is valid only when the surface emission is enough to achieve net SEEY that equals to one. When $\delta_{max}$ is low, incident primary electrons are not sufficiently accelerated to induce SEE and accumulate enough positive surface charges to form the SEEA. A transition to negative surface charge density occurs for reduced surface emission due to lower $\delta_{max}$. When $\delta_{max}$ approaches 1, which is true for some metallic materials, no SEEA can be formed regardless of $E_y$. This explains why a considerable difference with theory is observed only when $\delta_{max}$ is low in Figure 4. It has to be pointed out that the threshold SEEY coefficients $\delta_{max}$, $A_{max}$, or $E_y$ above which SEEA can be formed is rather complicated and no analytical expression is available. Here the SEEY parameters are particularly chosen to reveal the theory failure with less emissive surface. In the general case with dielectric surface, surface emissivity is strong and the discrepancy between simulation and SEEA theory is not remarkable.

As indicated by Equation (20), surface charge density is proportional to the applied electric field $E_y$. The SEEA initiation criterion is explained as follows. SEEA is formed only when the applied field is above certain level. This is because electrons emitted at CTJ by field emission carries low energy, which is comparable with the electrode material Fermi level. These low-energy electrons, if not sufficiently accelerated by the parallel field when arriving at the dielectric field, will accumulate negative surface charges with SEEY smaller than one. The accumulated negative surface charges will repel the following CTJ electrons instead of attracting them, such that a SEEA cannot be developed. An expression of the threshold electric field above which SEEA can develop has been calculated by a previous work [41], which has no analytical expression and is related to the angular distribution of field emission electrons, dielectric SEEY curve, and CTJ emission location. Generally, the threshold field is lower with larger $\delta_{max}$, smaller $A_{max}$ and higher CTJ emission location. In the present simulation setup, this threshold field is approximately $2\times10^6$ V/m, which is 0.4 times the adopted $E_y$ in simulation. The linear relation predicted by Equation (20) is well supported by the simulation results, shown in Figure 5.

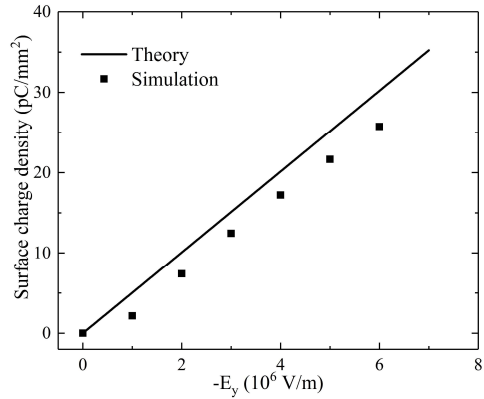

Figure 5. Comparison of simulated surface charge density with SEEA analytical prediction for varied applied field $E_y$. The theory curve is based on Equation (20).

## 4. Comparison with PIC model

In this section, the kinetic model is compared with the existing 2D2V PIC model, with the same parameter setup as listed in section 2, except with a 5mm length in the $y$ direction in the PIC model. More detailed descriptions of the adopted PIC model are presented in previous literatures [27, 32, 41]. The pros and cons of the kinetic model relative to the PIC model is discussed, aiming for an improved understanding of the model choices in future flashover modeling studies.

The 2D real-time positions of simulated macroparticles during SEEA development is shown in Figure 6. The SEEA is initiated near the CTJ due to field emission and develops towards the anode, until covering the whole insulator surface when reaching saturation. The 2D PIC model provides vivid descriptions of how the SEEA evolves whereas the present 1D2V model only studies one specific point in $y$ direction. Note that even after the SEEA reaches saturation, the vertical electron distribution is not perfectly uniform along the $y$ direction, and intense particle noises persist, as shown by the electron "spikes" of varied sizes above the insulator surface in Figure 6. A tiny slice of insulator near CTJ can carry negative surface charges, as the electrons colliding on that position are not sufficiently accelerated by $E_y$.

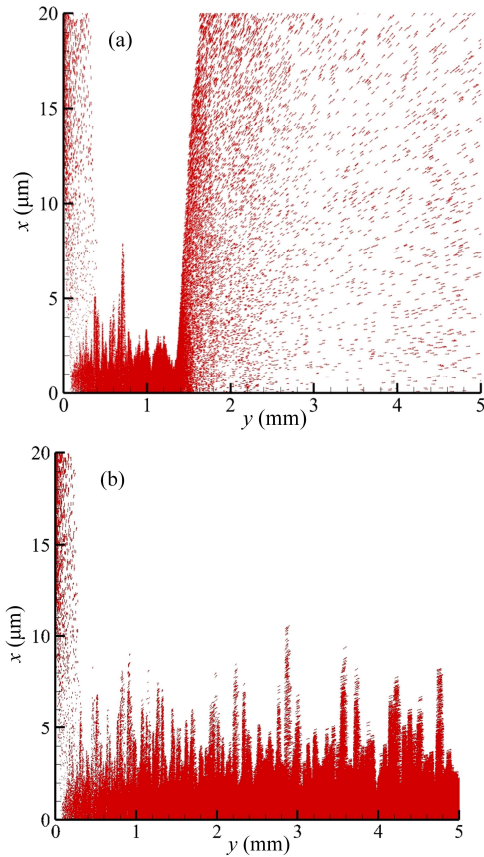

Figure 6. Electron position distribution in PIC model during SEEA development. (a) SEEA expansion. (b) SEEA saturation.

The reason for not adopting the 2D2V kinetic model in the present work is mainly due to limits on computational resources. In PIC model, adding dimensions requires updating the macroparticle coordinate setup. The PIC simulation speed mainly depends on the number of macroparticles. If the macroparticle number is fixed, the computation time approximately scales linearly with the number of extra dimensions, though in reality more microparticles are generally needed for simulation with higher spatial dimensions, in order to reveal the detailed kinetic effects. For kinetic model, the number of VDF matrix element grows exponentially with the number of dimensions, and more dimensions usually require a compromising decrease of grid resolution. For kinetic model with higher dimensions, parallelization of VDF matrix using e.g. MPI or OpenMP is needed, which is convenient to implement thanks to the adopted explicit numerical scheme.

A comparison of the surface charge density and incident primary electron flux towards the insulator surface given by the two simulation models is made and is shown in Figure 7. Analytical prediction of $\sigma_e$ is given by Equation (20), and the flux density prediction is $\Gamma_{pe} = \sigma_e/(et_{ef})$. Note that the time required for kinetic and PIC curves in Figure 7 to reach saturation are completely different and should not be confused. For kinetic model, the time to reach convergence is the time to fully charge the 1D insulator element. For the 2D PIC model, the time depends on the length of insulator and is equal to $d_l/v_{seea}$. Here $d_l$ is the insulator length (5mm) and $v_{seea}$ is the SEEA propagation velocity, measured to be of $10^7$ m/s by Anderson [68], which is consistent with the simulation. The comparison mainly focuses on the signal noise levels after reaching saturation, and the converged values given by the two simulation models.

The PIC model traces show significantly larger signal noises than traces from the kinetic model, particularly the $\Gamma_{pe}$ trace. The noise-free trace is one of the major advantages of using kinetic simulation approach. PIC model yields higher converged $\sigma_e$ and $\Gamma_{pe}$ values than the SEEA theory, and the kinetic model yields lower values than the SEEA theory. Discrepancies are lower for $\sigma_e$, with the kinetic model results closer to the theory prediction, as has been discussed in section 3. The larger $\sigma_e$ given by PIC model is likely related to the 2D effects. Averaging over the $y$ direction in PIC model causes derivation from the 1D model as well as the analytical prediction, due to the fact that SEEA is not perfectly uniform in $y$ direction, particularly near the CTJ, shown in Figure 6. The near-CTJ charging dynamics is sensitive to the choice of CTJ emission height and emission angle [69]. Discrepancies with theory predictions are also linked with the fact that the theory does not consider the space-charge effects and only analyzes single-particle trajectory under the applied field. The local field distortion caused by near-surface electrons can alter the electron trajectories. The discrepancies between simulation and theory are larger for primary electron flux as $\Gamma_{pe} \propto \sigma_e^2$ [70].

The reason for evaluating the surface charge density $\sigma_e$ above is that $\sigma_e$ is crucial to determine the flashover voltage. Though the present simulation model does not include the outgassing and breakdown stage of flashover, $\sigma_e$ can provide an estimation of the flashover threshold. It was shown that the local desorbed neutral pressure near the insulator surface is proportional to the square of $\sigma_e$ [70]. A higher desorbed neutral pressure will enhance ionization and decrease the breakdown voltage. By combining the linear relation between $\sigma_e$ and applied parallel electric field, and the empirical breakdown voltage formula of given gas species, Townsend first coefficient, pressure, the flashover voltage can be calculated. A next step plan is to upgrade the kinetic simulation model to include the breakdown stage of flashover and validate the obtained flashover voltage against the theory prediction, to be further discussed in section 5.

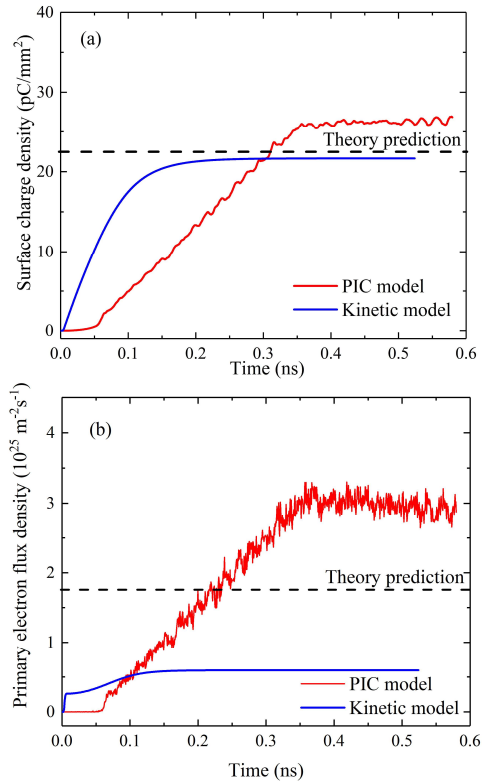

Figure 7. Comparison of surface charge density and primary electron flux density time traces given by the PIC model and kinetic model, with the same default simulation input parameters. Predictions given by SEEA analytical expression are marked by the dashed lines.

The comparison of kinetic and PIC simulation results suggests that the strong fluctuations of physical quantities such as electron fluxes and surface charge density observed in PIC simulations are due to numerical noises and do not represent realistic waves/instabilities.

## 5. Discussions

In the present simulation, only the electron VDF is simulated during the SEEA stage, without outgassing and the subsequent surface plasma discharge. The main motivation of the present work is to introduce and validate the implementation of the continuum, kinetic model in surface flashover simulation. In future code upgrades, the stages after SEEA will be included. A schematic of the surface flashover model updated from Figure 1, which includes the outgassing due to electron collision with the insulator surface and the ions created by electron-neutral ionization collision, is shown in Figure 8. Firstly, SEEA electrons collide on insulator surface, releasing previously adsorbed gas (desorption). The desorbed neutrals transport away from the insulator, and establish a local high-pressure region near the insulator. A discharge plasma is formed in the high-pressure region due to electron-neutral collisions and eventually leads to a breakdown. There exists a variety of challenges when simulating the outgassing and surface discharge processes in the PIC model.

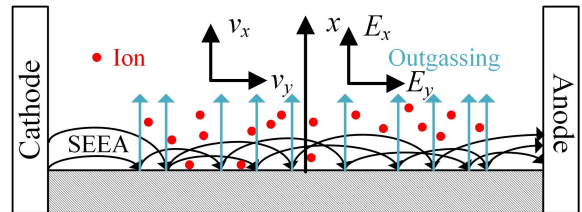

Figure 8. Schematic of the surface flashover model including outgassing and electron-neutral ionization collision.

First of all, the time scale of realistic outgassing process is significantly longer than the plasma discharge and SEEA. Typical local pressure near insulator can be above several Torr [16, 71], before the final plasma discharge, requiring outgassing time scale of order of 100 ns based on experimental observations [72]. This demands simplification in the PIC models such as presetting initial pressure or using greater outgassing rate to acceleration the simulation [16, 27]. This technical difficulty also exists the kinetic simulation.

The second difficulty is the drastically different particle density between SEEA stage and surface plasma discharge. As local pressure grows up due to outgassing, plasma density keeps increasing, so as the required number of microparticles. Though it is possible to adopt dynamics particle weight and adjust the weight in real-time simulation [73, 74], simulation speed during surface plasma discharge stage is in general well below the SEEA stage. The advantage of kinetic code, on the contrary, is that its simulation speed is not sensitive to the varying plasma density and only scales linearly with the number of particle species, which is doubled in discharge stage by introducing positive ion species.

In addition, the treatment of plasma-neutral collision is not straightforward in PIC model due to the strongly nonuniform neutral distribution, and special care is needed when performing null collision [75]. In contrast, plasma-neutral collision is realized in kinetic model by simply adding a BGK collision operator with specific collision frequency, which is proportional to the neutral density at each grid point and is easily calculated with existing collision cross-section data. Including outgassing and surface plasma discharge will hence not significantly reduce the kinetic simulation speed.

The choice of desorbed neutral transport model in the surface plasma discharge stage is also crucial. Direct simulation Monte Carlo (DSMC) method simulates the movement of individual neutral macroparticles, which are averaged over the grid point to obtain the near-surface pressure distribution. This method is compatible with the PIC model and was adopted in previous multipactor simulation with gas desorption [76]. Note that the flashover plasma is essentially a low temperature, partially ionized plasma, such that the neutral density is several orders of magnitude higher than the plasma density. The weight of neutral particle hence must be well above the weight of plasma macroparticles. In more recent surface flashover modeling [16], fluid model based on desorbed neutral diffusion equation was adopted. The analytical solution of neutral density profiles significantly reduces the required computational resources. In future kinetic modeling, the outgassing can be described by the desorbed neutral diffusion equation with a source boundary condition, prescribed by the real-time primary electron flux density and the gas desorption rate.

## 6. Conclusions

The 1D2V continuum and kinetic simulation model for surface flashover in vacuum is presented to reproduce the SEEA process. The simulation updates the electron velocity distribution function matrix by executing advections, electron-electron Coulomb collision, and field emission source at cathode triple junction. Secondary electron emission is implemented via the 2D matrix of SEE yield and VDF boundary condition, while updating the surface charge density simultaneously. The obtained surface charge density and flux density are consistent with the SEEA analytical prediction and the existing 2D PIC model, while the kinetic simulation provides noise-free data compared with the PIC model. A comparison between the kinetic and PIC simulation results suggests that the previously observed strong fluctuations in the PIC models are due to numerical noises instead of waves/instabilities. The simulation code can be upgraded to include the subsequent outgassing and surface plasma discharge stages in flashover, and is expected to show only limited reduction of simulation speed. The continuum and kinetic simulation is here validated as an alternative and effective numerical modeling approach for future surface flashover studies.

## Acknowledgements

This research was conducted under the auspices of the National Natural Science Foundation of China (NSFC) under Grants No. 51827809, 51707148 and No. 12175176. This work was supported in part by the Swiss National Science Foundation.